\documentclass[twoside,11pt]{article}

\usepackage[frozencache,cachedir=minted-cache]{minted}
\usepackage{obs_study_style}

\heading{}{2023}{}{}{}{Taehyeon Koo, Youjin Lee, Dylan S. Small, and Zijian Guo}

\ShortHeadings{RobustIV and controlfunctionIV}{Koo, Lee, Small, and Guo}
\firstpageno{1}

\usepackage{amsgen,amsmath,amsthm,amsfonts,amstext,amsbsy,amsopn,amssymb,latexsym}
\usepackage{bbm}
\usepackage[usenames]{color}
\usepackage{array}  
\usepackage{natbib}
\usepackage{tikz} 
\usetikzlibrary{arrows,intersections}
\usepackage{hyperref}
\usepackage{graphicx}  
\usepackage{amssymb}
\usepackage{float}
\usepackage{adjustbox} 
\usepackage{enumitem}

\usepackage{comment}
\includecomment{versionA}
\usepackage{xcolor} 
\definecolor{LightGray}{gray}{0.90}
\usepackage{bm}
\usepackage{comment}

 \usepackage[linesnumbered,ruled]{algorithm2e}
   
\let\emptyset\varnothing
\usepackage{multirow}

\newcommand\norm[1]{\left\lVert#1\right\rVert}
\newcommand{\hhat}{\widehat}
\newcommand{\E}{\mathbb{E}}
\newcommand{\I}{\mathbb{I}}

\newcommand{\Cov}{\text{Cov}}
\newcommand{\tr}{^{\mkern-1.5mu\mathsf{T}}}
\newcommand{\code}{\texttt}

\newcommand{\yes}{$\checkmark$}
\newcommand{\no}{ $\text{\sffamily X}$}

\newtheorem{Condition}{Condition}
\DefineVerbatimEnvironment{Sinput}{Verbatim}{fontshape=sl}
\DefineVerbatimEnvironment{Soutput}{Verbatim}{}
\DefineVerbatimEnvironment{Scode}{Verbatim}{fontshape=sl}

\DefineVerbatimEnvironment{Code}{Verbatim}{}
\DefineVerbatimEnvironment{CodeInput}{Verbatim}{fontshape=sl}
\DefineVerbatimEnvironment{CodeOutput}{Verbatim}{}
\newenvironment{CodeChunk}{}{}

\usepackage[normalem]{ulem}               
\newcommand\redout{\bgroup\markoverwith
{\textcolor{red}{\rule[0.5ex]{2pt}{0.8pt}}}\ULon}

\newcommand\Zijian[1]{{\color{magenta}Zijian: ``#1''}}

\title{\code{RobustIV} and \code{controlfunctionIV}: Causal Inference for Linear and Nonlinear Models with Invalid Instrumental Variables}
\author{\name Taehyeon Koo \email tk587@stat.rutgers.edu\\
       \addr Department of Statistics\\
       Rutgers University\\
       Piscataway, NJ 08854
       \AND
       \name Youjin Lee \email youjin\_lee@brown.edu\\
       \addr Department of Biostatistics\\
       Brown University\\
       Providence, RI 02912 
       \AND
       \name Dylan S. Small \email dsmall@wharton.upenn.edu\\
       \addr Department of Statistics\\
        The Wharton School, University of Pennsylvania\\
        Philadelphia, PA 19104
       \AND
       \name Zijian Guo \email zijguo@stat.rutgers.edu\\
       \addr Department of Statistics\\
       Rutgers University\\
       Piscataway, NJ 08854}
\date{\today}

\begin{document}

\maketitle
\begin{abstract}%
We present R software packages \code{RobustIV} and \code{controlfunctionIV} for causal inference with possibly invalid instrumental variables. \code{RobustIV} focuses on the linear outcome model. It implements the two-stage hard thresholding method to select valid instrumental variables from a set of candidate instrumental variables and make inferences for the causal effect in both low- and high-dimensional settings. Furthermore, \code{RobustIV} implements the high-dimensional endogeneity test and the searching and sampling method, a uniformly valid inference method robust to errors in instrumental variable selection. \code{controlfunctionIV} considers the nonlinear outcome model and makes inferences about the causal effect based on the control function method. Our packages are demonstrated using two publicly available economic data sets together with applications to the Framingham Heart Study.
\end{abstract}

\begin{keywords}
  Instrumental variable selection, confidence interval, nonlinear outcome model, control function, maximum clique
\end{keywords}

\section{Introduction}

A common problem in making causal inferences from observational studies is that there may be unmeasured confounders.  
The instrumental variable (IV) method is one of the most useful methods to estimate the causal effect when there might exist unmeasured confounding. The validity of IV methods relies on that the constructed IVs satisfy the following three assumptions simultaneously \citep[e.g.]{wooldridge2010}: conditioning the measured covariates,
\\
(A1) the IVs are associated with the treatment;\\
(A2) the IVs are independent with the unmeasured confounders;\\
(A3) the IVs have no direct effect on the outcome.

The main challenge of applying IV-based methods in practice is that the proposed IVs might not satisfy  the above assumptions (A1)-(A3). For example, in studying the causal effect of education on earning, the proximity of school \citep{AK1991,card1999} has been used as an instrumental variable. However, this instrument might be related to other factors, such as socioeconomic status, which could affect one’s earnings. Also,
there might be other advantages due to the proximity; for instance, people living close to college could be more likely to be exposed to vocational programs linked to colleges. So, the instrument could have a direct effect on earnings. In addition, the problem of IVs not satisfying assumptions (A1) to (A3) is a fundamental problem in Mendelian Randomization (MR), whose
goal is to estimate the causal effect of exposure on the disease by using genetic variants as instruments. These genetic variants might violate assumptions (A2) and (A3) due to pleiotropic effects \citep*{bowden2015,bowden2016,kang2016}.

This paper presents the R packages \code{RobustIV} and \code{controlfunctionIV},  implementing robust causal inference approaches proposed in \citet*{guo2018,guokang2018, guo2021,guo2016,li2020}.  The implemented inference methods choose the valid IVs among a set of candidate IVs that may violate the assumptions (A2) and (A3). The proposed methods target both linear and nonlinear causal effects. We also include the algorithm implementation for settings with high-dimensional covariates and IVs.

In the package \code{RobustIV}, we implement robust and high-dimensional IV algorithms for models assuming a constant and linear treatment effect. We implement the Two Stage Hard Thresholding (\texttt{TSHT}) proposed in \citet{guokang2018}, which selects valid IVs based on a voting method. The selected IVs are then used to infer the linear treatment effect. Additionally, \code{RobustIV} implements uniformly valid confidence intervals proposed in \citet{guo2021}, which guarantees valid coverage even if there are errors in selecting valid IVs. \code{RobustIV} also contains the high-dimensional endogeneity test proposed in \citet{guo2018}, generalizing the Durbin-Wu-Hausman test \citep{durbin1954,wu1973,hausman1978}.

On the other hand, we implement several control function methods in the package \code{controlfunctionIV} to infer causal effects under nonlinear outcome models. We apply the control function method for the continuous outcome variable by showing it as the two-stage least squares (TSLS) estimator with an augmented set of IVs \citep{guo2016}. We further follow \citet{guo2016} to test the validity of the augmented set of IVs and construct the pretest estimator by comparing the control function estimator and the TSLS estimator.  
We implement the probit control function method for the binary outcome and make inferences for the conditional average treatment effect (CATE) with possibly invalid IVs.
Moreover, the \code{controlfunctionIV} package implements the \code{SpotIV} method proposed in  
\citet{li2020} for the semi-parametric outcome model with possibly invalid IVs.

In \code{R}, there are well-developed IV methods when all IVs are assumed to satisfy the assumptions (A1) to (A3), such as \code{AER} by \citet{Kleiber2008} and \code{ivmodel} by \citet*{kang2020}. The main difference in our packages \code{RobustIV} and \code{controlfunctionIV} is that we allow for invalid IVs and leverage the multiple IVs to learn the validity of the candidate IVs. We shall mention other R packages implementing causal inference approaches with possibly invalid IVs: \code{sisvive} implemented the method proposed in \citet{kang2016} to estimate the treatment effect under the majority rule;
\code{CIIV} by \citet*{windmeijer2021} considered the causal inference approaches with possibly invalid IVs for the low-dimensional linear outcome model. In contrast, our packages \code{RobustIV} and \code{controlfunctionIV} are designed under a broader framework by allowing for linear and nonlinear outcome models with low- and high-dimensional IVs and covariates. 

Moreover, our package \code{RobustIV} provides a uniformly valid confidence interval robust to the errors in separating valid and invalid IVs, using the searching and sampling method proposed in \citet{guo2021}. \citet{guo2021} constructs uniformly valid confidence intervals by searching for a range of treatment effect values that lead to sufficiently many valid instruments. {There have been other uniformly valid inference and estimation methods developed for causal inference with possibly invalid IVs, such as \citet*{kang2020two} and \citet*{sun2023semi}. By specifying the minimum number of valid IVs instead of selection,  \citet{kang2020two} proposed a method for constructing a robust confidence interval for the causal effect by taking the union of well-known confidence intervals created from all models with the minimum number of valid IVs. \citet{sun2023semi} introduced the union causal model and defined the minimum number of valid IVs under the union model. By using the unconditional moment restrictions with interaction terms, the causal effect can be identified without ex-ante knowledge about which are valid IVs, and estimated via G-estimation.}

The GitHub repository at \url{https://github.com/bluosun/MR-GENIUS} implemented the \code{MR Genius} method  \citep*{tchetgen2021}, generalizing the method in \citet{lewbel2012hetero} and leveraging the heteroscedastic regression errors in the treatment model to identify the causal parameter. The R package \code{TSCI} implemented the two-stage curvature identification method proposed in \citet{guo2022tsci}, which leveraged machine learning methods to capture the nonlinearity in the treatment and identify the treatment effect with possibly invalid instruments. In contrast, our packages use different identification conditions than \citet{tchetgen2021,lewbel2012hetero,guo2022tsci}.

The paper is organized as follows. In Section \ref{section 2}, we review the methods {TSHT} and {Searching and Sampling} under the linear outcome model and demonstrate the usage of \code{TSHT} and \code{SearchingSampling} in \code{RobustIV} by analyzing economics data sets from \citet{AK1991}; in Section \ref{section endogeneity test}, we review the endogeneity test in high dimensions with possibly invalid IVs under the same model as Section \ref{section 2}, and give a demonstration of the usage of \code{endo.test} in \code{RobustIV} with simulated data. In Section \ref{section 3}, we discuss the inference approaches for the nonlinear outcome models implemented in \code{controlfunctionIV} and illustrate the usage of \code{controlfunctionIV} by analyzing {Mroz} data. In Section \ref{section 4}, we demonstrate our packages with an MR application to analyze the data from Framingham Heart Study (FHS). 
\paragraph{Notation.} 
 Let $\mathbb{R}^{p}$ be the set of real numbers with dimension $p$. For any vector $v \in \mathbb{R}^p$, $v_j$ denotes its $j$th element, $v_{-j}$ denotes whole $v$ except for $j$-th index, and  $\|v\|_0$ denotes the number of non-zero elements in $v$. For any $n \times p$ matrix
$M$, denote the $(i, j)$ entry by $M_{ij}$, the $i$th row by $M_{i\cdot}$, the $j$th column by $M_{\cdot j}$, and the transpose of $M$ by $M\tr$; also, $M_{IJ}$ denotes the submatrix of $M$ consisting of rows specified by the
set $I\subset \{1, . . . , n\}$ and columns specified by the set $J \subset \{1, . . . , p\}$, $M_{I\cdot}$ denotes the submatrix of $M$ consisting of rows indexed by the set I and all columns, and $M_{\cdot J}$ denotes the submatrix of M consisting of columns specified by the set $J$ and all rows. ${\rm I}_p$ denotes $p \times p$ identity matrix. $\I$ denotes the indicator function. $\Phi$ denotes the CDF of the standard normal distribution. For a sequence of random variable $X_n$, we use $X_n \overset{d}{\rightarrow} X$ to denote that $X_n$ converges to X in distribution.

\section{Linear outcome models}\label{section 2}
Throughout the paper, we consider $n$ i.i.d. observations. For $1\leq i\leq n,$ let $Y_{i} \in \mathbb{R}$, $D_{i} \in \mathbb{R},$ $Z_{i\cdot} \in \mathbb{R}^{p_z},$ and $X_{i\cdot} \in \mathbb{R}^{p_x}$ denote the outcome, the treatment, the instruments, and the baseline covariates, respectively.  
This section reviews the robust instrumental variable approaches in 
\citet{guokang2018, guo2021}, which are implemented in the \code{RobustIV} package. We demonstrate the usage of these functions in \code{RobustIV} in Sections \ref{sec: TSHT and SS}.

\subsection{Model assumption} \label{linear model assumption}

We assume the following outcome model with possibly invalid IVs \citep{small2007,kang2016,guokang2018,windmeijer2021}
\begin{align} 
     Y_{i} = D_{i}\beta+Z_{i\cdot}\tr \pi+X_{i\cdot}\tr \phi+\epsilon_{i}, \quad &\mathbb{E}[\epsilon_{i}Z_{i\cdot}]=0, \  \mathbb{E}[\epsilon_{i} X_{i\cdot}]=0  \label{linear eq}. 
\end{align}
 This is the linear structural model in econometrics \citep{wooldridge2010}. Here, we aim to estimate the constant causal effect $\beta\in \mathbb{R}$. 
If $D_{i}$ is correlated with $\epsilon_i$ in the model \eqref{linear eq}, we say it is an endogenous variable, and we cannot use popular estimators such as the OLS estimator. We also assume the linear association model for the treatment
\begin{align}
    D_{i} = Z_{i\cdot}\tr \gamma + X_{i\cdot}\tr \psi + \delta_{i}, \quad  &\mathbb{E}[\delta_{i}Z_{i\cdot}]=0, \ \mathbb{E}[\delta_{i}X_{i\cdot}]=0.\label{endo eq} 
\end{align}
As a remark, the errors in  \eqref{linear eq} and \eqref{endo eq}  are allowed to be heteroscedastic. In \eqref{linear eq} and \eqref{endo eq}, $\pi_j =0 $ if $j$-th IV satisfies the exclusion restriction conditions (A2) and (A3), and $\gamma_j \neq 0$ if it satisfies the strong IV assumption (A1). By assumptions in \eqref{linear eq} and \eqref{endo eq}, $\sigma_{12}=\Cov\left[\epsilon_{i},\delta_{i}|Z_{i\cdot},X_{i\cdot}\right]$ is the endogeneity parameter. We present the endogeneity test method under the violation of (A2) and (A3) in Section \ref{section endogeneity test}.

We discuss the causal interpretation of the above model \eqref{linear eq} using potential outcomes framework \citep{small2007,kang2016}. Let $Y_{i}^{(d,z)}$
be the potential outcome if individual $i$ were to receive the treatment $d$ and the instruments $z$. 
For two possible values of the treatment $d',d$ and instruments $z',z$, if we assume the following potential outcomes model
\begin{align}\label{causal eq}
Y_{i}^{(d',z')}-Y_{i}^{(d,z)}= (d'-d)\beta + (z'-z)\tr\kappa, \quad \mathbb{E}[Y_{i}^{(0,0)}|Z_{i\cdot},X_{i\cdot}]=X_{i\cdot}\tr \phi + Z_{i\cdot}\tr \eta,
\end{align}
and define $\pi=\kappa+\eta$, and $\epsilon_i=Y_{i}^{(0,0)}-\mathbb{E}[Y_{i}^{(0,0)}|Z_{i\cdot},X_{i\cdot}]$, we
obtain the model \eqref{linear eq}. {Here, $\eta$ and $\kappa$ are parameters that represent the violation of (A2) and (A3) respectively.}

By combining \eqref{linear eq} and \eqref{endo eq}, we obtain the reduced form models of $Y$ and $D$ as
\begin{align} 
    Y_{i} &= Z_{i\cdot}\tr \Gamma + X_{i\cdot}\tr \Psi + \xi_{i} , \quad \mathbb{E}[\xi_{i}Z_{i\cdot}]=0, \mathbb{E}[\xi_{i}X_{i\cdot}]=0, \label{reduced1} \\
    D_{i} &= Z_{i\cdot}\tr \gamma + X_{i\cdot}\tr \psi + \delta_{i} , \quad \mathbb{E}[\delta_{i}Z_{i\cdot}]=0, \mathbb{E}[\delta_{i}X_{i\cdot}]=0.  \label{reduced2}
\end{align}
Here, $\Gamma = \beta\gamma +\pi$, $\Psi = \beta\psi+\phi$ are reduced form parameters and $\xi_{i}=\beta\delta_{i}+\epsilon_{i}$ 
is the reduced form error term.


We introduce identifiability conditions for models \eqref{reduced1} and \eqref{reduced2}. 
Let $\mathcal{S}$ be the set of relevant IVs, i.e., $\mathcal{S}=\{1 \leq j \leq p_z: \gamma_{j} \neq 0 \}$ and $\mathcal{V}$ be the set of relevant and valid IVs, i.e.,
$\mathcal{V}=\{j\in \mathcal{S}: \pi_j=0\}$. The set $\mathcal{S}$ contains all candidate IVs that are strongly associated with the treatment. The set $\mathcal{V}$ is a subset of $\mathcal{S}$, which contains all candidate IVs satisfying all classical IV assumptions. The main challenge is that the set $\mathcal{V}$ is not known a priori in the data analysis. Additional identifiability conditions are needed for identifying the causal effect without any prior knowledge of $\mathcal{V}$. The majority rule is introduced to identify causal effects with invalid IVs \citep{bowden2016,kang2016}. 
\begin{Condition}[Majority Rule]\label{majority}
More than half of the relevant IVs are valid: $|\mathcal{V}| > |\mathcal{S}|/2$.
\end{Condition}
The following plurality rule is a weaker identification condition than the majority rule \citep*{Hartwig2017,guokang2018}. 
\begin{Condition}[Plurality Rule]\label{plurality}
The valid instruments form a plurality compared to the invalid instruments: $ |\mathcal{V}|>\max_{c\neq 0}|\{j \in \mathcal{S}: \pi_j/\gamma_j=c\}|.$
\end{Condition}

We present two inference methods for $\beta$ utilizing the majority and plurality: two stage hard thresholding and searching and sampling in Section \ref{section TSHT}. To present the methods, 
we consider the reduced form estimators $(\hhat{\Gamma}^{\intercal},\hhat{\gamma}^{\intercal})^{\intercal}$ satisfying 
\begin{align} \label{reduced form conv}
\sqrt{n}\left(\begin{pmatrix}
  \hhat{\Gamma} \\ \hhat{\gamma}
\end{pmatrix} - \begin{pmatrix}
  \Gamma \\ \gamma
\end{pmatrix}\right) \overset{d}{\rightarrow} N_{2p_z}\left(0_{2p_z}, \begin{pmatrix}
  \mathbf{V}^{\Gamma} & \mathbf{C} \\ \mathbf{C}\tr & \mathbf{V}^{\gamma}
\end{pmatrix}\right).
\end{align}
We use $\hhat{\mathbf{V}}^{\Gamma}, \hhat{\mathbf{C}},$ and $\hhat{\mathbf{V}}^{\gamma}$ to denote consistent estimators of asymptotic covariance matrix terms. In low dimensions, we estimate the reduced form $({\Gamma}^{\intercal},{\gamma}^{\intercal})^{\intercal}$  by the OLS estimator $(\hhat{\Gamma}^{\intercal},\hhat{\gamma}^{\intercal})^{\intercal}$ and estimate the variance covariance matrices by sandwich estimators; see the detailed construction in Section 2 of \citet{guo2021}. In high-dimensional settings, we can construct $(\hhat{\Gamma}^{\intercal},\hhat{\gamma}^{\intercal})^{\intercal}$ as the debiased Lasso estimator \citep*{belloni2011,javanmard2014,guokang2018}; see more details in Section 4.1 of \citet{guokang2018}.

\subsection{Two stage hard thresholding (TSHT) and Searching and Sampling}\label{section TSHT}
The TSHT consists of two steps: the first step is to screen out the weak IVs, and the second step is to screen out invalid IVs. Specifically, the first step of TSHT is to estimate the set $\mathcal{S}$ of relevant IVs  by 
$
    \hhat{\mathcal{S}}=\left\{ 1 \leq j \leq p_z : |\hhat{\gamma}_j| \geq \lambda_1 \sqrt{\hhat{\mathbf{V}}_{jj}^{\gamma}/n}\right\},
$
where $\lambda_1>0$ is a tuning parameter adjusting the testing multiplicity. 

The second thresholding step estimates the set $\mathcal{V}$ of valid instruments. Our main strategy is to assume that one IV is valid and evaluate whether the other IVs are valid from the point of view of that IV. Particularly, for $j \in \hhat{\mathcal{S}}$, we assume the $j$-th IV to be valid (i.e., $\pi_j=0$) and construct an estimator $\widehat{\pi}_{-j}$ of $\pi_{-j}$ using the equation $\pi_{-j}=\Gamma_{-j}-\beta^{[j]}\gamma_{-j}$ with $\beta^{[j]}=\Gamma_j/\gamma_j$, and get the standard error of the estimator. We test whether $\pi_{-j}=0$ by comparing $\widehat{\pi}_{-j}$ to a threshold, calculated as multiplying the standard error of $\widehat{\pi}_{-j}$ by a tuning parameter $\lambda_2$, which is a Bonferroni correction adjusting for testing multiplicity. Using the above test procedures, we construct a voting matrix $\tilde{\Pi} \in \mathbb{R}^{|\hhat{\mathcal{S}}|\times |\hhat{\mathcal{S}}|}$ where $\tilde{\Pi}_{j,k}= 1$ indicates that the $k$-th and $j$-th IVs agree with each other to be valid. Finally, we get a symmetric voting matrix $\hhat{\Pi}$ by setting $\hhat{\Pi}_{j,k}=\min\{\tilde{\Pi}_{j,k},\tilde{\Pi}_{k,j}\}$.

Once we get $\hhat{\Pi}$, we estimate $\mathcal{V}$ by two options. Let $\text{VM}_k$ denote the number of votes that the $k^{\rm th}$ IV, with $k \in \hhat{\mathcal{S}}$, received from other candidates of IVs. First, we define $\hhat{\mathcal{V}}$ by the set of IVs that receive a majority and a plurality of votes \citep{guokang2018}
\begin{align}
    \hhat{\mathcal{V}}^\code{MP}:=\{k \in \hhat{\mathcal{S}}:\text{VM}_k > |\hhat{\mathcal{S}}|/2\}\cup\{k \in \hhat{\mathcal{S}}:\text{VM}_k=\max_{l\in\hhat{\mathcal{S}}}\text{VM}_l\}.
    \label{vhat mp}
\end{align}
The next method is to estimate $\mathcal{V}$ by the maximum clique method. We can generate a graph $\mathcal{G}$ with indexes belonging to $\hhat{\mathcal{S}}$ and the adjacency matrix as $\hhat{\Pi}$. That is, the indexes $j,k\in \hhat{\mathcal{S}}$ are connected if and only if $\hhat{\Pi}_{j,k}=1$. Then as suggested in \citet{windmeijer2021}, we can estimate $\hhat{\mathcal{V}}^{\code{MC}}$ as the maximum clique of the graph $\mathcal{G}$, which is the largest fully connected sub-graph of $\mathcal{G}$ \citep{Csardi2006}. Note that there might be several maximum cliques. In this case, each maximum clique forms an estimator of ${\mathcal{V}}$ and our proposal reports several causal effect estimators based on each maximum clique. 

We further illustrate the definitions of $\hhat{\mathcal{V}}^{\code{MP} }$ and $\hhat{\mathcal{V}}^{\code{MC}}$ using the following example. Consider $p_z = 8$ with $\{z_1, z_2, z_3, z_4\}$ being valid and $\{z_5, z_6, z_7\}$ being invalid with the same invalidity level, and $z_8$ being invalid IV with a different invalidity level. The left side of Table \ref{tab:voting matrix table} corresponds to an ideal setting where the valid and invalid IVs are well separated and the valid IVs $\{z_1, z_2, z_3, z_4\}$ only vote for each other. In this case, $\hhat{\mathcal{V}}^{\code{MC}}=\hhat{\mathcal{V}}^\code{MP}=\{z_1, z_2, z_3, z_4\}$. On the right side of Table \ref{tab:voting matrix table}, we consider the setting that the invalidity level of $z_5$ might be mild and the IV $z_5$ receives the votes from three valid IVs $\{z_2, z_3, z_4\}$. In this case, $\hhat{\mathcal{V}}^\code{MP}=\{z_2, z_3, z_4, z_5\}$. In contrast, there are two maximum cliques $\{z_1, z_2, z_3, z_4\}$ and $\{z_2, z_3, z_4, z_5\}$ and   $\hhat{\mathcal{V}}^\code{MC}$ can be either of these two. 


\begin{table}[t]
\centering
 \parbox{.46\linewidth}{
        
  \begin{tabular}{r|llllllllll}
           & $z_1$ & $z_2$& $z_3$&$z_4$&$z_5$&$z_6$&$z_7$&$z_8$\\
            \hline
        $z_1$ &$\yes$&$\yes$&$\yes$&$\yes$&$\no$&$\no$&$\no$&$\no$\\
         
         $z_2$&$\yes$&$\yes$&$\yes$&$\yes$&$\no$&$\no$&$\no$&$\no$\\
         $z_3$&$\yes$&$\yes$&$\yes$&$\yes$&$\no$&$\no$&$\no$&$\no$\\
         $z_4$&$\yes$&$\yes$&$\yes$&$\yes$&{$\no$}&$\no$&$\no$&$\no$\\
         $z_5$&$\no$&{$\no$}&$\no$&{$\no$}&$\yes$&$\yes$&$\yes$&$\no$\\
         $z_6$&$\no$&$\no$&$\no$&$\no$&$\yes$&$\yes$&$\yes$&$\no$\\
         $z_7$&$\no$&$\no$&$\no$&$\no$&$\yes$&$\yes$&$\yes$&$\no$\\
         $z_8$&$\no$&$\no$&$\no$&$\no$&$\no$&$\no$&$\no$&$\yes$\\ 
         \hline 
         Votes&4&4&4&4&3&3&3&1\\
        \end{tabular}
    }
\hfill
\parbox{.46\linewidth}{
       
  \begin{tabular}{r|llllllllll}
           & $z_1$ & $z_2$& $z_3$&$z_4$&$z_5$&$z_6$&$z_7$&$z_8$\\
            \hline
        $z_1$ &$\yes$&$\yes$&$\yes$&$\yes$&$\no$&$\no$&$\no$&$\no$\\
         
         $z_2$&$\yes$&$\yes$&$\yes$&$\yes$&$\yes$&$\no$&$\no$&$\no$\\
         $z_3$&$\yes$&$\yes$&$\yes$&$\yes$&$\yes$&$\no$&$\no$&$\no$\\
         $z_4$&$\yes$&$\yes$&$\yes$&$\yes$&{$\yes$}&$\no$&$\no$&$\no$\\
         $z_5$&$\no$&{$\yes$}&$\yes$&{$\yes$}&$\yes$&$\yes$&$\yes$&$\no$\\
         $z_6$&$\no$&$\no$&$\no$&$\no$&$\yes$&$\yes$&$\yes$&$\no$\\
         $z_7$&$\no$&$\no$&$\no$&$\no$&$\yes$&$\yes$&$\yes$&$\no$\\
         $z_8$&$\no$&$\no$&$\no$&$\no$&$\no$&$\no$&$\no$&$\yes$\\ 
         \hline 
         Votes&4&5&5&5&6&3&3&1\\
        \end{tabular}
    }
 
    \caption{The left voting matrix $\hhat{\Pi}$ denotes that all valid IVs $\{z_1, z_2, z_3, z_4\}$ vote each other but not any other invalid IV. The right voting matrix $\hhat{\Pi}$ denotes that the locally
invalid IV $z_5$ receives votes from valid IVs $\{z_2, z_3, z_4\}$ and invalid IVs $\{z_6, z_7\}$.}
\label{tab:voting matrix table}
\end{table}

Once we have $\hhat{\mathcal{V}}$, we can construct an efficient point estimator $\hhat{\beta}$ for $\beta$ in a low-dimensional setting via one-step iteration as follows. First, we construct an initial estimator $\tilde{\beta}=\frac{\hhat{\gamma}_{\hhat{\mathcal{V}}}\tr \tilde{A} \hhat{\Gamma}_{\hhat{\mathcal{V}}}}{\hhat{\gamma}_{\hhat{\mathcal{V}}}\tr \tilde{A} \hhat{\gamma}_{\hhat{\mathcal{V}}}},$ where $ \tilde{A}=\hhat{\Sigma}_{\hhat{\mathcal{V}},\hhat{\mathcal{V}}}-\hhat{\Sigma}_{\hhat{\mathcal{V}},\hhat{\mathcal{V}}^c}\hhat{\Sigma}_{\hhat{\mathcal{V}}^c,\hhat{\mathcal{V}}^c}^{-1}\hhat{\Sigma}_{\hhat{\mathcal{V}}^c,\hhat{\mathcal{V}}}$, $\hhat{\Sigma}=\frac{1}{n}\sum_{i=1}^nW_{i\cdot}W_{i\cdot}\tr$, and $W_{i\cdot}=(Z_{i\cdot}\tr, X_{i\cdot}\tr)\tr$.
    
 Next, we get a point estimator $\hhat{\beta}$ by one-step iteration \citep{holland1977robust}
\begin{align} \label{betalow}
    \hhat{\beta}=\frac{\hhat{\gamma}_{\hhat{\mathcal{V}}}\tr \hhat{A} \hhat{\Gamma}_{\hhat{\mathcal{V}}}}{\hhat{\gamma}_{\hhat{\mathcal{V}}}\tr \hhat{A} \hhat{\gamma}_{\hhat{\mathcal{V}}}}, \quad \text{where} \quad \hhat{A}= [(\hhat{\mathbf{V}}^\Gamma-2\tilde{\beta} \hhat{\mathbf{C}} +\tilde{\beta}^2\hhat{\mathbf{V}}^\gamma)_{\hhat{\mathcal{V}},\hhat{\mathcal{V}}}]^{-1}.
\end{align}
Finally, the $1-\alpha$ confidence interval for $\beta$ is
\begin{align} \label{ci low}
    (\hhat{\beta}-z_{1-\alpha/2}\hhat{\rm SE},\hhat{\beta}+z_{1-\alpha/2}\hhat{\rm SE}) \quad \text{where} \quad \hhat{\rm SE}= \sqrt{\frac{\hhat{\gamma}_{\hhat{\mathcal{V}}}\tr\hhat{A} (\hhat{\mathbf{V}}^\Gamma-2\hhat{\beta} \hhat{\mathbf{C}} +\hhat{\beta}^2\hhat{\mathbf{V}}^\gamma)_{\hhat{\mathcal{V}},\hhat{\mathcal{V}}}\hhat{A}\hhat{\gamma}_{\hhat{\mathcal{V}}}}{n(\hhat{\gamma}_{\hhat{\mathcal{V}}}\tr\hhat{A} \hhat{\gamma}_{\hhat{\mathcal{V}}})^2}}.
\end{align}
As a remark, $\hhat{\mathbf{V}}^\Gamma, \hhat{\mathbf{V}}^\gamma,$ and $\hhat{\mathbf{C}}$ are heteroscedasticity-robust covariance estimators and hence \eqref{ci low} is also robust to heteroscedastic errors in a low-dimensional setting. In a high-dimensional setting, we set $\hhat{A}={\rm I}$ in \eqref{betalow}, and $\hhat{\mathbf{V}}^\Gamma, \hhat{\mathbf{V}}^\gamma,$ and $\hhat{\mathbf{C}}$ are constructed under the homoscedastic error assumptions; see more details in \citet{guokang2018}. 

We now review the searching and sampling method proposed in \citet{guo2021}, which provides uniformly valid conference intervals even if there are errors in separating valid and invalid IVs. The right-hand side of Table \ref{tab:voting matrix table} illustrates an example of the invalid IVs not being separated from valid IVs in finite samples. In the following, we review the idea of searching and sampling under the majority rule and the more general method with the plurality rule can be found in \citet{guo2021}. 

Let $\alpha\in (0,1)$ denote the pre-specified significance level. Given $\beta \in \mathbb{R}$ and the reduced form estimator $\hhat{\Gamma}$ and $\hhat{\gamma}$, 
we estimate $\pi_j$ with $j \in \hhat{\mathcal{S}}$ by 
\begin{equation}
\hhat{\pi}_j(\beta)=(\hhat{\Gamma}_j-\beta\hhat{\gamma}_j)\I(|\hhat{\Gamma}_j-\beta\hhat{\gamma}_j|\geq \hhat{\rho}_j(\beta,\alpha)),
\label{eq: thresholding searching}
\end{equation}
where  $\hhat{\rho}_j(\beta,\alpha) = \Phi^{-1}\left(1-\frac{\alpha}{2|\hhat{\mathcal{S}}|}\right)\hhat{\text{SE}} (\hhat{\Gamma}_j-\beta\hhat{\gamma}_j)$ with $\hhat{\text{SE}} (\hhat{\Gamma}_j-\beta\hhat{\gamma}_j)$ denoting a consistent estimator of the standard error of $\hhat{\Gamma}_j-\beta\hhat{\gamma}_j$.
We search for the value of $\beta$ leading to enough valid IVs and construct the searching confidence interval as
\begin{align} \label{Searching ci}
     \text{CI}^{\text{search}}= \left\{\beta \in \mathbb{R}:\norm{\hhat{\pi}_{\hhat{\mathcal{S}}}(\beta)}_0<|\hhat{\mathcal{S}}|/2 \right\},
\end{align}
which collects all $\beta$ values such that more than half of IVs in $\widehat{\mathcal{S}}$ are selected as valid.

Based on the searching method, \citet{guo2021} proposed a sampling confidence interval, which retains the uniform coverage property and improves the precision of the confidence interval.
In particular, we sample 
\begin{align*} \label{sampling 1}
    \begin{pmatrix}
    \hhat{\Gamma}^{[m]} \\
    \hhat{\gamma}^{[m]}
    \end{pmatrix} \overset{iid}{\sim}
    N\left[
    \begin{pmatrix}
    \hhat{\Gamma} \\
    \hhat{\gamma}
    \end{pmatrix},
    \frac{1}{n}\begin{pmatrix}
    \hhat{\mathbf{V}}^{\Gamma} & \hhat{\mathbf{C}} \\
    \hhat{\mathbf{C}}\tr & \hhat{\mathbf{V}}^{\gamma}
    \end{pmatrix}
    \right], \quad \text{for}\quad 1\leq m\leq M.
\end{align*}

For $1\leq m \leq M$ and $j \in \hhat{\mathcal{S}}$, we modify \eqref{eq: thresholding searching} and define $$\hhat{\pi}_{j}^{[m]}(\beta,\lambda)=(\hhat{\Gamma}_j^{[m]}-\beta\hhat{\gamma}_j^{[m]})\I(|\hhat{\Gamma}_j^{[m]}-\beta\hhat{\gamma}_j^{[m]}|\geq \lambda \cdot \hhat{\rho}_j(\beta,\alpha))$$ with the shrinkage parameter $\lambda \asymp \left({\log n}/{M}\right)^{\frac{1}{2|\hhat{\mathcal{S}}|}}.$ A data-dependent way of choosing $\lambda$ can be found in Remark 3 of \citet{guo2021}. 
For each $1 \leq m \leq M$, we construct a searching interval $(\beta_{\min}^{[m]},\beta_{\max}^{[m]})$  where $$\beta_{\min}^{[m]} = \min_{\beta \in \mathcal{B}_\lambda^{[m]}}\beta\quad \text{and}\quad \beta_{\max}^{[m]} = \max_{\beta \in \mathcal{B}_\lambda^{[m]}} \beta$$  with $\mathcal{B}_\lambda^{[m]} = \left\{\beta \in \mathbb{R}:\norm{\hhat{\pi}_{\hhat{\mathcal{S}}}^{[m]}(\beta,\lambda)}_0<|\hhat{\mathcal{S}}|/2 \right\}$.
Then the sampling CI is defined as
\begin{equation} \label{sampling ci}
    \text{CI}^{\text{sample}}=\left(\min_{m\in \mathcal{M}}\beta_{\min}^{[m]},\max_{m\in \mathcal{M}}\beta_{\max}^{[m]}\right).
\end{equation}
with $\mathcal{M}=\{1 \leq m \leq M : (\beta_{\min}^{[m]},\beta_{\max}^{[m]})\neq\emptyset\}.$ The sampling confidence intervals in general improve the precision of the searching confidence intervals. But both intervals can provide uniformly valid coverage robust to the errors in separating valid and invalid IVs. 
\subsection{\code{TSHT} and \code{SearchingSampling} Usage}
\label{sec: TSHT and SS}
In this section, we introduce usages of the R functions \code{TSHT} and \code{SearchingSampling} in the \code{RobustIV} package, functions for the methods in Section \ref{section TSHT}, with the data used in \citet{AK1991}. \citet{AK1991} studied the causal effect of the years of education (\code{EDUC}) on the log weekly earnings (\code{LWKLYWGE}). Following \citet{AK1991}, we take 30 interactions (\code{QTR120}-\code{QTR129}, \code{QTR220}-\code{QTR229}, \code{QTR320}-\code{QTR329}) between three quarter-of-birth dummies (\code{QTR1}-\code{QTR3}) and ten year-of-birth dummies (\code{YR20}-\code{YR29}) as the instruments $Z$. For example, \code{QTR120} is element-wise product of \code{QTR1} and \code{YR20}. Here, the quarter-of-birth dummies are the indicators of whether the observed person was born in the first, second, and third quarter of the year respectively, and the year-of-birth dummies are indicators of which year the subject was born from 1940 to 1949 respectively. We also include the following baseline covariates $X$: 9 year-of-birth dummies (\code{YR20}-\code{YR28}), a race dummy (\code{RACE}), a marital status dummy (\code{MARRIED}), a dummy for residence in an SMSA (\code{SMSA}), and eight region-of-residence dummies (\code{NEWENG, MIDATL, ENOCENT, WNOCENT, SOATL, ESOCENT, WSOCENT, MT}). 
We first apply the function \code{TSHT}.

\begin{minted}[bgcolor=LightGray, fontsize=\footnotesize]{R}
R> Y <- as.vector(LWKLYWGE); D <- as.vector(EDUC)
R> Z <- sapply(paste0("QTR", c(seq(120,129), seq(220,229), seq(320,329))),
               function(x){get(x)})
R> X <- cbind(sapply(paste0("YR",seq(20,28)),function(x){get(x)}),RACE, MARRIED,  
             SMSA, NEWENG, MIDATL, ENOCENT, WNOCENT, SOATL, ESOCENT, WSOCENT, MT)
R> pz <- ncol(Z)
R> out.TSHT <- TSHT(Y=Y,D=D,Z=Z,X=X, 
                    tuning.1st = sqrt(2.01*log(pz)), tuning.2nd = sqrt(2.01*log(pz)))
R> summary(out.TSHT)
 betaHat Std.Error CI(2.5%) CI(97.5%) Valid IVs                                       
 0.0874  0.019     0.0502   0.1247    QTR120 QTR121 QTR122 QTR220 QTR222 QTR227 QTR322
_ _ _ _ _ _ _ _ _ _ _ _ _ _ _ _ _ _ _ _ _ _ _ _ _ _ _ _ _ _ 
Detected invalid IVs: QTR126 QTR226
\end{minted}
Here, \code{tuning.1st} and \code{tuning.2nd} are tuning parameters $\lambda_1$ and $\lambda_2$ used for the thresholds to get $\hhat{\mathcal{S}}$ and $\hhat{\mathcal{V}}$ in Section \ref{section TSHT} respectively. The default values for these parameters are $\sqrt{\log n}$ in the low-dimensional setting. However, in theory, any value above $\sqrt{2\log p}$ and diverging to infinity would suffice. Since the data has 486926 observations, we choose the tuning parameters as $\sqrt{2.01\log p_z}$ to avoid too conservative threshold levels due to the huge sample. Once \code{TSHT} is implemented, we can call \code{summary} to see the outputs of \code{TSHT} including the point estimator, its standard error, confidence interval, and valid IVs as we discussed in Section \ref{section TSHT}.

The above result shows that \code{TSHT} selected \code{QTR120}, \code{QTR121}, \code{QTR122}, \code{QTR220}, \code{QTR222}, \code{QTR227}, and \code{QTR322} as valid IVs. Thus, valid IVs are interactions with the first quarter of birth and dummies representing births in 1940, 1941, and 1942, interactions with the second quarter of birth and dummies representing births in 1940, 1942, and 1947, and finally the interaction between the third quarter-of-birth and dummy representing births in 1942. On the other hand, it is reported that \code{QTR126} and \code{QTR226} are invalid IVs. That is, interactions with the first and the second quarter of birth and dummy representing births in 1946 are relevant but invalid IVs. {The remaining IVs have been screened out of the first-stage selection as individually weak IVs.}

The detection of invalid IVs implies that using whole $Z$ as valid IVs can cause the estimate to be biased. In \citet{AK1991}, the TSLS estimate by using whole $Z$ as valid IVs is 0.0393. In contrast, our procedure is more robust to the existence of possibly invalid IVs, giving the causal estimate as 0.0874.
Our $95\%$ confidence interval is above zero, indicating a positive effect of education on earning.

In addition to the above output, the class object \code{TSHT} has other values that are not reported by \code{summary}, for example, whether the majority rule is satisfied or not, and the voting matrix to construct $\hhat{\mathcal{V}}$ in Section \ref{section TSHT}. These can be checked by directly calling \code{TSHT}.

As discussed in Section \ref{section TSHT}, there are different voting options to get $\hhat{\mathcal{V}}$, where the default option \code{voting = 'MaxClique'} stands for $\hhat{\mathcal{V}}^{\code{MC}}$ and \code{voting = 'MP'} stands for $\hhat{\mathcal{V}}^{\code{MP}}$ in \eqref{vhat mp}. If there are several maximum cliques, \code{summary} returns results corresponding to each maximum clique. Furthermore, since the default argument for which estimator to use is \code{method = 'OLS'}, one can choose other estimators by \code{method = 'DeLasso'} for the debiased Lasso estimator with \code{SIHR} R package \citep*{rakshit_cai_guo_2021} and \code{method = 'Fast.DeLasso'} for the fast computation of the debiased Lasso estimator \citep{javanmard2014}. The above methods are useful in a high-dimensional setting. 

 Next, we implement the uniformly valid confidence intervals by calling the function \code{SearchingSampling}. We start with the searching CI defined in \eqref{Searching ci} with the argument \code{Sampling = FALSE}. 
 \begin{minted}[bgcolor=LightGray, fontsize=\footnotesize]{R}
R> out1 = SearchingSampling(Y=Y, D=D, Z=Z, X=X, Sampling=FALSE, 
                   tuning.1st = sqrt(2.01*log(pz)), tuning.2nd = sqrt(2.01*log(pz)))
R> summary(out1)
Confidence Interval for Causal Effect: [-0.0964,0.2274]
\end{minted}
With the default argument \code{Sampling = TRUE}, one can use the following code to implement the more efficient sampling CI in \eqref{sampling ci}.
 \begin{minted}[bgcolor=LightGray, fontsize=\footnotesize]{R}
R> set.seed(1)
R> out.SS = SearchingSampling(Y=Y, D=D, Z=Z, X=X,
                      tuning.1st = sqrt(2.01*log(pz)), tuning.2nd = sqrt(2.01*log(pz)))
R> summary(out.SS)
Confidence Interval for Causal Effect: [0.0135,0.1775]
\end{minted}

The \code{SearchingSampling} confidence intervals are generally wider than that of the \code{TSHT} since they are robust to the IV selection errors.
The function \code{summary} displays confidence interval for $\beta$, which are discussed in Section \ref{section TSHT}. As in \code{TSHT}, one can use the argument \code{method} to employ the high-dimensional debiased estimators instead of OLS.

\section{Endogeneity test}\label{section endogeneity test}
In this section, we present a method for testing endogeneity under the same models as in Section \ref{section 2} with the null hypothesis $H_0: \sigma_{12}=0$, specifically tailored for high-dimensional settings, considering the potential violation of assumptions (A2) and (A3). In Section \ref{sec: endotest method}, we review the endogeneity test method presented in \citet{guo2018}; in Section \ref{sec: endotest}, we give a demonstration of the usage of \code{endo.test} in \code{RobustIV}.
\subsection{Endogeneity test in high dimensions with possibly invalid IVs} \label{sec: endotest method}
We review the high-dimensional endogeneity test proposed in \citet{guo2018}. We focus on the homoscedastic error setting by writing 
$\Theta_{11}=\text{Var}[\xi_{i}|Z_{i\cdot},X_{i\cdot}]$, $\Theta_{22}=\text{Var}[\delta_i |Z_{i\cdot}, X_{i\cdot}]$, and $\Theta_{12}=\text{Cov}[\xi_{i},\delta_{i}|Z_{i\cdot},X_{i\cdot}]$ for the reduced form models \eqref{reduced1} and \eqref{reduced2}. With the same estimators from TSHT in the high-dimensional settings in Section \ref{section TSHT}, we can estimate the covariance $\sigma_{12}$ by $\hhat{\sigma}_{12}=\hhat{\Theta}_{12}-\hhat{\beta}\hhat{\Theta}_{22}$ where
$\hhat{\beta}=\frac{\sum_{j \in \hhat{\mathcal{V}}}\hhat{\gamma}_{j}\hhat{\Gamma}_{j}}{\sum_{j \in \hhat{\mathcal{V}}}\hhat{\gamma}_{j}^2}$. Here, $\hhat{\Theta}_{12}$ and $\hhat{\Theta}_{22}$ are consistent estimators of the reduced form covariance ${\Theta}_{22}$ and $\Theta_{12}$, and $\hhat{\mathcal{V}}$ is an estimator of $\mathcal{V}$ as in Section \ref{section TSHT}. We establish the asymptotic normality of $\hhat{\sigma}_{12}-{\sigma}_{12}$ in \citet{guo2018} and propose a testing procedure for $H_0: {\sigma}_{12}=0$. 

\subsection{\code{endo.test} Usage}
\label{sec: endotest}
In the following, we show the usage of \code{endo.test} in the \code{RobustIV} package, a function for the endogeneity test in high dimension with a simulated example. The corresponding model and method are presented in Section \ref{section TSHT} and Section \ref{section endogeneity test} respectively. 
We consider the models \eqref{linear eq} and \eqref{endo eq} and set $p_z=600$ with only the first 10 IVs being relevant. Among these 10 IVs,  the first 3 IVs are invalid but the remaining IVs are valid. Moreover, we set $\text{Corr}\left(\epsilon_i,\delta_i\right)=0.8$, which indicates a level of endogeneity. The function $\code{endo.test}$ generates a class object with same arguments in \code{TSHT}. The class object from \code{endo.test} can be used by calling \code{summary} function, which enable us to see a brief result of \code{ento.test}.
  
\begin{minted}[bgcolor=LightGray, fontsize=\footnotesize]{R}
R> set.seed(5)
R> n = 500; L = 600; s = 3; k = 10; px = 10; epsilonSigma = matrix(c(1,0.8,0.8,1),2,2)
R> beta = 1; gamma = c(rep(1,k),rep(0,L-k))
R> phi = (1/px)*seq(1,px)+0.5; psi = (1/px)*seq(1,px)+1
R> Z = matrix(rnorm(n*L),n,L); X = matrix(rnorm(n*px),n,px);
R> epsilon = MASS::mvrnorm(n,rep(0,2),epsilonSigma)
R> D = 0.5 + Z %*% gamma + X %*% psi + epsilon[,1]
R> Y = -0.5 + Z %*% c(rep(1,s),rep(0,L-s)) + D * beta + X %*% phi + epsilon[,2]
R> endo.test.model <- endo.test(Y,D,Z,X, invalid = TRUE)
R> summary(endo.test.model)
 P-value Test        Valid IVs            
 0       H0 rejected Z4 Z5 Z6 Z7 Z8 Z9 Z10
_ _ _ _ _ _ _ _ _ _ _ _ _ _ _ _ _ _ _ _ _ _ _ _ _ _ _ _ _ _ 
Detected invalid IVs: Z1 Z2 Z3 
\end{minted}
When we call \code{summary} function, $p$-value, it reports the test result with significance level $\alpha$ (default \code{alpha = 0.05}), the valid IVs, and detected invalid IVs. \code{H0 rejected} means that the treatment is endogenous, otherwise not. Since we set \code{invalid = TRUE}, \code{ento.test} allows some of IVs to be invalid and conducts the endogeneity test with the selected $\hhat{\mathcal{V}}$ defined in Section \ref{section TSHT}. With \code{invalid = FALSE}, the function assumes that all IVs are valid. 
As in Section \ref{sec: TSHT and SS}, one can use \code{method} argument to employ other estimators both in the low and high dimensions.


\section{Nonlinear outcome models} 
\label{section 3}
This section reviews the control function IV methods 
\citep{guo2016,li2020} implemented in the \code{controlfunctionIV} package, whose usage is demonstrated in Sections \ref{sec: cf} and \ref{sec: Probit.cf}.

\subsection{Control function and pretest estimators} \label{section cf and pretest}
We consider the following nonlinear outcome and treatment models:
\begin{align}
    Y_{i}&= \mathbf{G}(D_i)\tr\boldsymbol{\beta} + X_{i\cdot}\tr\phi + u_i, \ \E[u_iZ_{i\cdot}]=\E[u_iX_{i\cdot}]=0, \label{additive y}\\
    D_{i}&=\mathbf{H}(Z_{i\cdot})\tr\boldsymbol{\gamma}+X_{i\cdot}\tr\mathbf{\psi} +v_{i}, \ \E[v_iZ_{i\cdot}]=\E[v_iX_{i\cdot}]=0 \label{additive d}, 
\end{align}
where $\mathbf{G}(D_i)=(D_i,g_2(D_i),...,g_k(D_i))\tr$,  $\mathbf{H}(Z_{i\cdot})=(Z_{i\cdot},h_{2}(Z_{i\cdot}),...,h_{k}(Z_{i\cdot}))\tr$ with $\{g_j(\cdot)\}_{2\leq j\leq k}$ and $\{h_j(\cdot)\}_{2\leq j\leq k}$ denoting the known nonlinear transformations. Under the models \eqref{additive y} and \eqref{additive d}, the IVs are assumed to be valid and the causal effect of increasing the value of $D$ from $d_2$ to $d_1$ is defined as
$
     \mathbf{G}(d_1)\tr\boldsymbol{\beta}-\mathbf{G}(d_2)\tr\boldsymbol{\beta}.
$

The control function (CF) method is a two-stage procedure. In the first stage, regress $D$ on $\mathbf{H}(Z)$ and $X$, and obtain the predicted value $\widehat{D}$ and its associated residual $\hhat{v}=D-\hhat{D}$. In the second stage, we use $\hhat{v}$ as the proxies for the unmeasured confounders and regress $Y$ on $\mathbf{G}(D)$, $X$, and $\hhat{v}$. We use $\hhat{\beta}^{\text{CF}}$ to denote the estimated regression coefficient corresponding to $D$. \citet{guo2016} showed that $\hhat{\beta}^{\text{CF}}$ is equivalent to the TSLS estimator with the augmented set of IVs. Even if all IVs satisfy the classical assumptions (A1)-(A3), there is no guarantee of the validity of the augmented IVs generated by the CF estimator. \citet{guo2016} applied the Hausman test to assess the validity of the augmented set of IVs generated by the CF estimator. The test statistic is defined as  
\begin{align} \label{hausman}
    H(\hhat{\beta}^{\text{CF}},\hhat{\beta}^{\text{TSLS}})=(\hhat{\beta}^{\text{CF}}-\hhat{\beta}^{\text{TSLS}})\tr [\Cov(\hhat{\beta}^{\text{TSLS}})-\Cov(\hhat{\beta}^{\text{CF}})]^{-}(\hhat{\beta}^{\text{CF}}-\hhat{\beta}^{\text{TSLS}}),
\end{align}
where $\hhat{\beta}^{\text{TSLS}}$ is the two stage least square estimator, $\Cov(\hhat{\beta}^{\text{TSLS}})$ and $\Cov(\hhat{\beta}^{\text{CF}})$ are the covariance matrices of $\hhat{\beta}^{\text{TSLS}}$ and $\hhat{\beta}^{\text{CF}}$, and $A^{-}$ denote the Moore-Penrose pseudoinverse. 
If the $p$-value $P\left(\chi_{1}^2 \geq H(\hhat{\beta}^{\text{CF}},\hhat{\beta}^{\text{TSLS}})\right)$ is less than $\alpha = 0.05$, then we define the level $\alpha$ pretest estimator $\hhat{\beta}^{\text{Pretest}}$ as $\hhat{\beta}^{\text{TSLS}}$; otherwise, $\hhat{\beta}^{\text{CF}}$ defined above \citep{guo2016}.  A code implementation of these methods is in Section \ref{sec: cf}.
\subsection{\code{cf} and \code{pretest} Usage} 
\label{sec: cf}
 In this section, we introduce usages of \code{cf} and \code{pretest} in the package \code{controlfunctionIV} using the {Mroz} data set from \citet{wooldridge2010}. The {Mroz} data was introduced in \citet{mroz1987} and then used in various works of literature including \citet{wooldridge2010}, which has $n = 428$ individuals after removing the data with \code{NA}. Following \citet{wooldridge2010}, we estimate the causal effect of education on the log earnings of married working women. The data is available in the \code{Wooldridge} package.

Here, the outcome $Y$ is log earnings (\code{lwage}), and the exposure $D$ is years of schooling (\code{educ}). Moreover, there are other variables such as the father's education (\code{fatheduc}), the mother's education (\code{motheduc}), the husband's education (\code{huseduc}), actual labor market experience (\code{exper}), its square (\code{expersq}), and the women's age (\code{age}). 

Following Example 5.3 in \citet{wooldridge2010}, we assume \code{motheduc}, \code{fatheduc}, and \code{huseduc} to be valid IVs, denoted as $Z_i = (Z_{i1},Z_{i2},Z_{i3})\tr$; we use and \code{exper}, \code{expersq}, and \code{age} as baseline covariates, denoted as $X_i = (X_{i1},X_{i2},X_{i3})\tr.$ Also assume that the outcome and treatment models are \eqref{additive y} and \eqref{additive d} respectively with $\mathbf{G}(D_i)=(D_i,D_i^2)\tr$ and $\mathbf{H}(Z_{i\cdot})=(Z_{i1},Z_{i2},Z_{i3},Z_{i1}^2,Z_{i2}^2,Z_{i3}^2)\tr$. {In practice, it is important to note that model specification may vary depending on the context of the IV model being used. Additionally, expert knowledge, subject matter understanding, and careful consideration of the underlying causal structure are crucial for model specification. Even in the case where the correct models are not fully known, researchers can employ model selection procedures with IV methods, for example, \citet{andrews2001consistent}.}

We can implement the \code{cf} function by inputting a \code{formula} object, which has the same form as that of \code{ivreg} in \code{AER} package. The function \code{summary} gives us information on coefficients of the control function estimators, including the point estimator, its standard error, $t$ value, and $p$ value.
\begin{minted}[bgcolor=LightGray, fontsize=\footnotesize]{R}
R> library(wooldridge); library(controlfunctionIV); data(mroz); mroz <- na.exclude(mroz)
R> Y <- mroz[,"lwage"]; D <- mroz[,"educ"]
R> Z <- as.matrix(mroz[,c("motheduc","fatheduc","huseduc")])
R> X <- as.matrix(mroz[,c("exper","expersq","age")])
R> cf.model <- cf(Y~D+I(D^2)+X|Z+I(Z^2)+X)
R> summary(cf.model)
_ _ _ _ _ _ _ _ _ _ _ _ _ _ _ _ _ _ _ _ _ _ _ _ _ _ _ _ _ _ 
Coefficients of the control function estimators:

              Estimate  Std.Error t value Pr(>|t|)    
(Intercept)  1.2573907  0.7871438   1.597 0.055457 .  
D           -0.1434395  0.1102058   1.302 0.096884 .  
I(D^2)       0.0086426  0.0041004   2.108 0.017817 *  
Xexper       0.0438690  0.0131574   3.334 0.000465 ***
Xexpersq    -0.0008713  0.0003984   2.187 0.014631 *  
Xage        -0.0011636  0.0048634   0.239 0.405511    
---
Signif. codes:  0 ‘***’ 0.001 ‘**’ 0.01 ‘*’ 0.05 ‘.’ 0.1 ‘ ’ 1
\end{minted}

The following code infers the causal effect $\mathbf{G}(d_1)\tr\boldsymbol{\beta}-\mathbf{G}(d_2)\tr\boldsymbol{\beta}$ by changing the treatment level from $d_2$ to $d_1=d_2+1$. Since the second and third coefficients are related to $D$, we use the second and third index to get the causal effect and its standard error. 

\begin{minted}[bgcolor=LightGray, fontsize=\footnotesize]{R}
R> d2 = median(D); d1 = median(D)+1;
R> D.diff <- c(d1,d1^2)-c(d2,d2^2); CE <- (D.diff)%*%cf.model$coefficients[c(2,3)]
R> CE.sd <-sqrt(D.diff%*%cf.model$vcov[c(2,3),c(2,3)]%*%D.diff)
R> CE.ci <- c(CE-qnorm(0.975)*CE.sd,CE+qnorm(0.975)*CE.sd)
R> cmat <- cbind(CE,CE.sd,CE.ci[1],CE.ci[2])
R> colnames(cmat)<-c("Estimate","Std.Error","CI(2.5%)","CI(97.5%)"); rownames(cmat)<- "CE"
R>  print(cmat, digits = 4)
   Estimate Std.Error CI(2.5%) CI(97.5%)
CE  0.07263   0.02171  0.03007    0.1152
\end{minted}
The function 
\code{pretest} can be used to choose between the TSLS or the control function method. If we run \code{pretest} with the same argument above and call \code{summary}, it will output the following result:
\begin{minted}[bgcolor=LightGray, fontsize=\footnotesize]{R}
R> pretest.model <- pretest(Y~D+I(D^2)+X|Z+I(Z^2)+X)
R> summary(pretest.model)

Level 0.05 pretest estimator is control function estimator. 
_ _ _ _ _ _ _ _ _ _ _ _ _ _ _ _ _ _ _ _ _ _ _ _ _ _ _ _ _ _ 
Coefficients of the pretest estimators:

              Estimate  Std.Error t value Pr(>|t|)    
(Intercept)  1.2573907  0.7871438   1.597 0.055457 .  
D           -0.1434395  0.1102058   1.302 0.096884 .  
I(D^2)       0.0086426  0.0041004   2.108 0.017817 *  
Xexper       0.0438690  0.0131574   3.334 0.000465 ***
Xexpersq    -0.0008713  0.0003984   2.187 0.014631 *  
Xage        -0.0011636  0.0048634   0.239 0.405511    
---
Signif. codes:  0 ‘***’ 0.001 ‘**’ 0.01 ‘*’ 0.05 ‘.’ 0.1 ‘ ’ 1
\end{minted}
 The first section of the output of \code{summary} reports which estimator is chosen after the pretesting step. The second section lists brief information on coefficients of pretest estimators including the point estimator, its standard error, $t$ value, and $p$-value, similar to \code{cf}. Since the pretest estimator is the control function estimator, the second section of \code{summary} is the same as that of \code{summary(cf.model)}.

\subsection{Probit CF and SpotIV}
\label{section probit cf}
We now consider the binary outcome model and continuous treatment model, 
\begin{align}
    &\E\left[Y_{i}|D_i=d, W_{i\cdot}=w, u_i=u\right]=\I(d\beta+w\tr \kappa+u>0), \quad \text{and}\quad D_{i}=W_{i\cdot}\tr \gamma + v_{i},
    \label{probit model} 
\end{align}
where $W_{i\cdot}=(Z_{i\cdot}\tr, X_{i\cdot}\tr)\tr$, the errors $(u_i,v_i)^{\intercal}$ are bivariate normal random variables with zero means and independent of $W_{i\cdot}$, $\kappa = (\kappa_{z}\tr,\kappa_{x}\tr)\tr$ is the coefficient vector of the IVs and measured covariates, and $\gamma=(\gamma_{z}\tr,\gamma_{x}\tr)\tr$ is a parameter representing the association between $D_{i}$ and $W_{i\cdot}$. When $\kappa_z\neq 0,$ the instruments are invalid. 
Since $u_i$ and $v_i$ are bivariate normal, we write $u_i=\rho v_i+e_i.$ The model \eqref{probit model} implies $\E\left[Y_{i}|W_{i\cdot}, v_i\right]=\Phi(D_i\beta^*+W_{i\cdot}^{\intercal}\Gamma^*+\rho^*v_i)$
where $\beta^*=\beta/\sigma_e $, $\Gamma^*=\kappa/\sigma_e +\beta^*\cdot \gamma$, and $\rho^*=\rho/\sigma_e+\beta^*$ with $\sigma_e$ denoting the standard error of $e_i=u_i-\rho v_i.$ That is, the conditional outcome model of $Y_i$ given $W_{i,\cdot}$ and $v_i$ is a probit regression model. 

Our goal is to estimate the conditional average treatment effect (CATE) from $d_2$ to $d_1$
\begin{align}
    \text{CATE}(d_1,d_2|w)&:=\mathbb{E}[Y_{i}|D_i=d_1,W_{i\cdot}=w]-\mathbb{E}[Y_{i}|D_i=d_2,W_{i\cdot}=w]. \label{Probit cate def}
\end{align}

We first construct the OLS estimator $\hhat{\gamma}$ of $\gamma$. We compute its residual $\hhat{v}=D-W\hhat{\gamma}$ and define $\hhat{\Sigma}=\frac{1}{n}
\sum_{i=1}^nW_{i\cdot}W_{i\cdot}\tr$. We estimate $\mathcal{S}=\{1\leq j \leq p_z: (\gamma_z)_j \neq 0\}$ by
\begin{align} \label{probit shat}
    \hhat{\mathcal{S}}=\left \{ 1 \leq j \leq p_z : |\hhat{\gamma}_j| \geq \hhat{\sigma}_v\sqrt{2\{\hhat{\Sigma}^{-1}\}_{j,j}\log n/n}\right \}
\end{align}
with $\hhat{\sigma}_{v}^2=\sum_{i=1}^{n}\hhat{v}_i^2/n$. Next, as CF in Section \ref{section cf and pretest}, we use $\widehat{v}$ as the proxy for unmeasured confounders and implement the  probit regression $Y$ on $W$ and $\hhat{v}$. We use $\hhat{\Gamma}$ and $\hhat{\rho}$ to denote the probit regression coefficients of $W$ of $\hhat{v}$ respectively. We apply the majority rule and compute $\hhat{\beta}$ as the median of $(\hhat{\Gamma}_j/\hhat{\gamma}_j)_{j \in \hhat{\mathcal{S}}}.$ We then estimate  $\hhat{\kappa}=\hhat{\Gamma} - \hhat{\gamma}\hhat{\beta}$. Finally, we estimate CATE defined in \eqref{Probit cate def} by the partial mean method \citep*{newey1994kernel,mammen2012nonparametric}, 
\begin{align*}
    \frac{1}{n}\sum_{i=1}^n\left[\Phi(d_1\hhat{\beta}+w\tr\hhat\kappa+\hhat{v}_i\hhat\rho)\right]-\frac{1}{n}\sum_{i=1}^n\left[\Phi(d_2\hhat{\beta}+w\tr\hhat\kappa+\hhat{v}_i\hhat\rho)\right]
\end{align*}
and construct the confidence interval by bootstrap  \citep{li2020}. 

\citet{li2020} has proposed a more general methodology, named \code{SpotIV}, to conduct robust causal inference with possibly invalid IVs. The model considered in \citet{li2020} includes the probit outcome model in \eqref{probit model} as a special case. In particular, \citet{li2020} replaced the known probit transformation in \eqref{probit model} with the more general non-parametric function, which is possibly unknown. Moreover, \citet{li2020} allows some instruments to be correlated with the unmeasured confounders $u_i$ in the outcome model. 
\subsection{\code{Probit.cf} Usage}
\label{sec: Probit.cf}
In this section, we look at the usage of \code{Probit.cf}, which is designed for the binary outcome with unmeasured confounders and possibly invalid IVs. For illustration, we use the {Mroz} data in Section \ref{sec: cf} and define the binary outcome variable $Y_0$ to take the value 1 if the continuous outcome $Y$ is greater than the median of $Y$ and 0 otherwise. We use the same treatment variable $D$ as in the \code{cf} example. Contrary to the \code{cf} example, we set the candidates of IVs $Z$ as \code{motheduc}, \code{fatheduc}, \code{huseduc}, \code{exper}, and \code{expersq}, and assume that we have covariates $X$ as \code{age}. 

We implement the \texttt{Probit.cf} function to estimate the CATE by increasing the treatment value from the median of $D$ to the median plus one. We can call \code{summary} to see the result of \code{Probit.cf}.
The function \code{summary} provides information on the valid IVs $\hhat{\mathcal{V}}$, the point estimator, standard error, and $95\%$ confidence interval for $\beta$ in \eqref{probit model}, and the point estimator, the standard error, and $95\%$ confidence interval of CATE. 
\begin{minted}[bgcolor=LightGray, fontsize=\footnotesize]{R}
R> Z <- as.matrix(mroz[,c("motheduc","fatheduc","huseduc","exper","expersq")])
R> Y0 <- as.numeric((Y>median(Y)))
R> d2 = median(D); d1 = d2+1; w0 = apply(cbind(Z,X)[which(D == d2),], 2, mean)
R> Probit.model <- Probit.cf(Y0,D,Z,X,d1 = d1,d2 = d2,w0 = w0)
R> summary(Probit.model)
     Estimate Std.Error CI(2.5%) CI(97.5%) Valid IVs                
Beta 0.2119   0.092     0.0316   0.3922    motheduc fatheduc huseduc
CATE 0.0844   0.033     0.0198   0.1489    motheduc fatheduc huseduc
_ _ _ _ _ _ _ _ _ _ _ _ _ _ _ _ _ _ _ _ _ _ _ _ _ _ _ _ _ _ 
No invalid IV is detected
\end{minted}
With the option \code{invalid = TRUE}, we allow invalid IVs and choose the valid IVs among all provided IVs. If one wants to assume all IVs are valid, one can set \code{invalid = FALSE}. 

\section{Application to Framingham Heart Study}
\label{section 4}

We analyze the Framingham Heart Study (FHS) data and illustrate our package using genetic variants as IVs. The FHS is an ongoing cohort study of participants from the town of Framingham, Massachusetts, that has grown over the years to include five cohorts with a total sample of over 15,000. The FHS, initiated in 1948, is among the most critical sources of data on cardiovascular epidemiology \citep*{sytkowski1990changes, kannel2000framingham, mahmood2014framingham}. Since the late 1980s, researchers across human health-related fields have used genetic factors underlying cardiovascular diseases and other disorders. Over the last two decades, DNA has been collected from blood samples and immortalized cell lines from members of the Original Cohort, the Offspring Cohort, and the Third Generation Cohort~\citep{govindaraju2008genetics}. Several large-scale genotyping projects and genome-wide linkage analysis have been conducted, and several other recent collaborative projects have completed thousands of SNP genotypes for candidate gene regions in subsets of FHS subjects with available DNA. 
The FHS has recently been used for Mendelian Randomization to determine causal relationships even in the presence of unmeasured confounding thanks to the availability of genotype and phenotype data~\citep{holmes2014causal, dalbeth2015mendelian, hughes2014mendelian}.
As candidate IVs, we will use genotype data from the FHS associated with the phenotype of interest and apply the proposed methods described above.


We apply the \code{RobustIV} package to investigate the effect of low-density lipoprotein (LDL-C) on globulin levels among individuals in the Framingham Heart Study (FHS) Offspring Cohort, as was studied in~\cite{kang2020two}. 
We use eight SNP genotypes (rs646776, rs693, rs2228671, rs2075650, rs4299376, rs3764261, rs12916, rs2000999) that are known to be significantly associated with LDL-C measured in mg/dL as candidate IVs \citep{kathiresan2007genome, ma2010genome,smith2014association}. See Table~\ref{tab:candidate} for details. The outcome of interest $Y_{i}$ is a continuous globulin level ($g/L$) and the exposure variable $D_{i}$ is the LDL-C level. Globulin is known to play a crucial role in liver function, clotting, and the immune system. We also use the age and sex of the subjects as covariates ${X}_{i\cdot}$.
The study includes $n=1445$ subjects, with an average globulin level of $27.27$ (SD: $3.74$) and an average LDL-C of $1.55$ (SD: 0.50). An average age is $35.58$ (SD: $9.74$) and $54.95\%$ are males. 
\begin{table}[H]
\centering
\resizebox{1.0\textwidth}{!}{\begin{tabular}{lllllll}
\hline
\multirow{2}{*}{$Z_{j}$} & \multirow{2}{*}{SNP} & \multirow{2}{*}{Position} & \multicolumn{2}{c}{\texttt{lm(D $\sim$ Z)}} & \multicolumn{2}{c}{\texttt{lm(Y $\sim$ Z)}} \\ 
& & & Estimate (Std. Error) & $t$-statistic ($p$-value) & Estimate (Std. Error) & $t$-statistic ($p$-value) \\
\hline
$Z_{1}$ & rs646776 & chr1:109275908  &  -5.160 (1.610) &  -3.205  (0.001) & -0.001 (0.170) & -0.007 (0.994)  \\
$Z_{2}$ & rs693 & chr2:21009323 &  -3.600  (1.286) & -2.799  (0.005)  &   0.318  (0.135) &  2.349  (0.019)    \\ 
$Z_{3}$ & rs2228671 & chr19:11100236 &  7.138   (2.029)  & 3.518 ($<$0.001) &  0.529    (0.214) &  2.474  (0.014)  \\
$Z_{4}$ & rs2075650 & chr19:44892362 & 8.451 (2.021) &  4.183 ($<$0.001) & 0.471    (0.213) &  2.208   (0.027)  \\
$Z_{5}$ & rs4299376 & chr2:43845437 &  3.847    (1.387)  & 2.773  (0.006)  & 0.110  (0.146) &  0.752(0.452)  \\
$Z_{6}$ & rs3764261 & chr16:56959412 &  3.651    (1.429)  & 2.555   (0.011)   &  0.275   (0.151)  & 1.829   (0.067)    \\
$Z_{7}$ & rs12916 & chr5:75360714 &  3.363 (1.365) &  2.463   (0.014)  & -0.195     (0.144) & -1.357    (0.175)  \\
$Z_{8}$ & rs2000999 & chr16:72074194 &   -2.961 (1.629) & -1.818   (0.069)    & -0.119  (0.172) & -0.691 (0.489)  \\
\hline
\end{tabular}}
\caption{\label{tab:candidate} Summary of the relationship between the single nucleotide polymorphisms (SNPs) and low-density lipoprotein. The point estimator, its standard error, $t$ value, and $p$-value are summary statistics from running a marginal regression model specified in the column title. Position refers to the position of the SNP in the chromosome, denoted as chr.}
\end{table}



By applying \texttt{endo.test}, we detect one invalid IV and observe the evidence for the existence of unmeasured confounders since the null hypothesis $H_{0}: \sigma_{12} = 0$ is rejected. 

\begin{minted}[bgcolor=LightGray, fontsize=\footnotesize]{R}
R> pz <- ncol(Z)
R> globulin.endo2 <- endo.test(Y,D,Z,X, invalid = TRUE,
        tuning.1st = sqrt(2.01*log(pz)), tuning.2nd = sqrt(2.01*log(pz)))
R> summary(globulin.endo2)
P-value Test        Valid IVs              
 0.0091  H0 rejected Z.1 Z.3 Z.4 Z.5 Z.6 Z.8
_ _ _ _ _ _ _ _ _ _ _ _ _ _ _ _ _ _ _ _ _ _ _ _ _ _ _ _ _ _ 
Detected invalid IVs: Z.2 
\end{minted}

Next, we implement \texttt{TSHT} with the default method of \texttt{"OLS"} under the low-dimensional setting. Again, the same invalid IV is detected and the confidence interval is above zero, indicating a positive effect of LDL on the glucose level.  

\begin{minted}[bgcolor=LightGray, fontsize=\footnotesize]{R}
R> pz <- ncol(Z)
R> TSHT2 <- TSHT(Y, D, Z, X,  
                tuning.1st = sqrt(2.01*log(pz)), tuning.2nd = sqrt(2.01*log(pz)))
R> summary(TSHT2)
 betaHat Std.Error CI(2.5%) CI(97.5%) Valid IVs              
 0.0529  0.0146    0.0243   0.0814    Z.1 Z.3 Z.4 Z.5 Z.6 Z.8
_ _ _ _ _ _ _ _ _ _ _ _ _ _ _ _ _ _ _ _ _ _ _ _ _ _ _ _ _ _ 
Detected invalid IVs: Z.2 
\end{minted}


We also constructed the confidence interval using the searching method, which provides robustness to the IV selection errors.
\begin{minted}[bgcolor=LightGray, fontsize=\footnotesize]{R}
R> SS1 <- SearchingSampling(Y, D, Z, X, tuning.1st = sqrt(2.01*log(pz)), 
                tuning.2nd = sqrt(2.01*log(pz)), Sampling = FALSE)
R> summary(SS1)
Confidence Interval for Causal Effect: [-0.2427,0.1894]
\end{minted}
We further implement the sampling method, which leads to a shorter uniformly valid CI than the searching method.
\begin{minted}[bgcolor=LightGray, fontsize=\footnotesize]{R}
R> SS2 <- SearchingSampling(Y, D, Z, X, tuning.1st = sqrt(2.01*log(pz)), 
                tuning.2nd = sqrt(2.01*log(pz)), Sampling = TRUE)
R> summary(SS2)
Confidence Interval for Causal Effect: [-0.0521,0.1259]
\end{minted}

In the following, we study nonlinear causal relationships using the \code{controlfunctionIV} package. 
\citet*{burgess2014instrumental} investigated a nonlinear causal relationship between BMI and diverse cardiovascular risk factors. Here we examine BMI's possibly nonlinear causal effect on the insulin level. Among $n=3733$ subjects, we excluded 618 subjects with missing information on insulin level, and 50 subjects whose insulin level is greater than $300\text{pmol/L}$ and whose BMI is greater than $45\text{kg/m}^{2}$. We use log-transformed insulin as the outcome of interest $Y_{i}$ measured at Exam 2. 
The exposure $D_{i}$ denotes the BMI measures at Exam 1. 
The covariates $X_{i \cdot}$ that we adjusted for are age and sex. As valid IVs $Z_{i \cdot}$, we propose using four SNP genotypes known to be significantly associated with obesity. In our analysis, we include \texttt{I(D\^{}2)} and \texttt{I(X\^{}2)} to account for quadratic effects of BMI, age, and sex on the outcome. We also include \texttt{I(Z\^{}2)} to account for possible quadratic effects of SNPs on the exposure. The result from the pretest estimator is as follows:

\begin{minted}[bgcolor=LightGray, fontsize=\footnotesize]{R}
R> insulin.pretest = pretest( Y ~ D + I(D^2) + X  + I(X^2) | Z + I(Z^2) + X + I(X^2))
R> summary(insulin.pretest)

Level 0.05 pretest estimator is control function estimator. 
_ _ _ _ _ _ _ _ _ _ _ _ _ _ _ _ _ _ _ _ _ _ _ _ _ _ _ _ _ _ 

Coefficients of Pretest Estimators:

              Estimate    Std.Err t value Pr(>|t|)    
(Intercept)  2.674e+00  6.006e-01   4.453 4.39e-06 ***
D            8.295e-02  2.828e-02   2.933 0.001690 ** 
I(D^2)      -7.784e-04  2.742e-04   2.839 0.002276 ** 
X1          -1.780e-02  7.816e-03   2.277 0.011427 *  
I(X1^2)      2.954e-04  8.852e-05   3.337 0.000428 ***
X2          -1.361e-01  5.654e-02   2.406 0.008087 ** 
---
Signif. codes:  
0 ‘***’ 0.001 ‘**’ 0.01 ‘*’ 0.05 ‘.’ 0.1 ‘ ’ 1
\end{minted}
The pretest estimator chooses the control function over the standard TSLS. The results also show that BMI has a positive linear effect on the outcome but a negative quadratic effect on the outcome. 

\section*{Acknowledgement} The research of T. Koo was supported in part by NIH grants R01GM140463 and R01LM013614. The research of D. Small was supported in part by NIH grant 5R01AG065276-02.The research of Z. Guo was partly supported by the NSF grants DMS 1811857 and 2015373 and NIH grants R01GM140463 and R01LM013614. Z. Guo is grateful to Dr. Frank Windmeijer for bringing up the maximum clique method.

The Framingham Heart Study is conducted and supported by the National Heart, Lung, and Blood Institute (NHLBI) in collaboration with Boston University (Contract No. N01-HC-25195, HHSN268201500001I, and 75N92019D00031). This manuscript was not prepared in collaboration with investigators of the Framingham Heart Study and does not necessarily reflect the opinions or views of the Framingham Heart Study, Boston University, or NHLBI.
Funding for SHARe Affymetrix genotyping was provided by NHLBI Contract N02-HL64278. SHARe Illumina genotyping was provided under an agreement between
Illumina and Boston University. Funding for Affymetrix genotyping of the FHS Omni cohorts was provided by Intramural NHLBI funds from Andrew D. Johnson and Christopher J. O’Donnell.
\vskip 0.2in
\bibliography{ref}

\end{document}